\documentclass[conference]{IEEEtran}

\usepackage[cmex10]{amsmath}
\usepackage{amsfonts}
\usepackage{mathrsfs}

\usepackage{amssymb}   
\usepackage{algorithm}
\usepackage{algorithmicx}
\usepackage{algpseudocode}
\usepackage{cite}      
\usepackage[dvips]{graphicx}
\usepackage{epstopdf}
\usepackage{psfrag}    
\usepackage{subfigure} 
\usepackage{url}       
\usepackage{stfloats}  
\usepackage{enumerate}
\usepackage{bm}
\usepackage{fancyhdr}

\usepackage{threeparttable}

\usepackage{booktabs}

\usepackage{caption}
\begin{document}

%
\newcommand\Pact {\mathscr{P}}
\title{RIS-Assisted Coordinated Multi-Point ISAC  for Low-Altitude Sensing Coverage} 


\author{\IEEEauthorblockN{Ying Zhang\IEEEauthorrefmark{1}\IEEEauthorrefmark{2}, Zeqi Hao\IEEEauthorrefmark{1}\IEEEauthorrefmark{2}, and
Tingting Zhang\IEEEauthorrefmark{1}\IEEEauthorrefmark{2}\IEEEauthorrefmark{3}}
\IEEEauthorblockA{\IEEEauthorrefmark{1}
School of Electronics and Information Engineering, \\
Harbin Institute of Technology, Shenzhen, P. R. China}
\IEEEauthorblockA{\IEEEauthorrefmark{2} Guangdong Provincial Key Laboratory of Space-Aerial Networking and Intelligent Sensing}
\IEEEauthorblockA{\IEEEauthorrefmark{3} Peng Cheng Laboratory, Shenzhen, P. R. China\\
Email: zhangying@stu.hit.edu.cn, 24S052020@stu.hit.edu.cn, zhangtt@hit.edu.cn
}
}

\maketitle

\begin{abstract}
The low-altitude economy (LAE) has emerged and developed in various fields, which has gained considerable interest. To ensure the security of LAE, it is essential to establish a proper sensing coverage scheme for monitoring the unauthorized targets. Introducing integrated sensing and communication (ISAC) into cellular networks is a promising solution that enables coordinated multiple base stations (BSs) to significantly enhance sensing performance and extend coverage. Meanwhile, deploying a reconfigurable intelligent surface (RIS) can mitigate signal blockages between BSs and low-altitude targets in urban areas. Therefore, this paper focuses on the low-altitude sensing coverage problem in RIS-assisted coordinated multi-point ISAC networks, where a RIS is employed to enable multiple BSs to sense a prescribed region while serving multiple communication users. A joint beamforming and phase shifts design is proposed to minimize the total transmit power while guaranteeing sensing signal-to-noise ratio and communication spectral efficiency. To tackle this non-convex optimization problem, an efficient algorithm is proposed by using the alternating optimization and semi-definite relaxation techniques. Numerical results demonstrate the superiority of our proposed scheme over the baseline schemes.

\end{abstract}


\begin{IEEEkeywords}
Integrated sensing and communication (ISAC), reconfigurable intelligent surface (RIS), sensing coverage,  beamforming, coordinated multi-point.
\end{IEEEkeywords}

%
\IEEEpeerreviewmaketitle

\section{Introduction}
The low-altitude economy (LAE) is an economic form centered on low-altitude flight activities of various manned and unmanned aerial vehicles (UAVs). It drives the development of related fields such as intelligent logistics, environment monitoring and emergency response, attracting tremendous attention from academia and industry \cite{Zhao2025}. However, the successful implementation of LAE requires full-time and full-area security. Therefore, it is need to provide ubiquitous sensing to detect unauthorized ``black flight'' drones and targets within a specific airspace \cite{Meng2024,Wu2021}.

The integrated sensing and communication (ISAC), a candidate technique for next-generation wireless networks, enables the implementation of both sensing and communication functions on a unified platform. Consequently, a promising solution for monitoring unauthorized low-altitude targets is integrating ISAC into cellular networks, where the cooperation of multiple base stations (BSs) can significantly expand coverage range and enhance the sensing performance by multi-dimensional target measurements and high spatial diversity \cite{Zhang2026}. Moreover, it can properly mitigate the air-ground interference in sensing and communication via implementing joint transmission at the central processors \cite{Tang2025,Jiang2025}.

Some studies have recently investigated cellular ISAC for low-altitude sensing coverage. In \cite{Li2024}, the transmit beamforming  was designed to concurrently sense a prescribed region while serving a group of communication users. The successive convex approximation technique and mesh grid approach were proposed to solve the non-convex optimization problem. Zhang \emph{et al}. in \cite{Zhang2025} jointly optimized the ISAC transmit and receive beamformers at the BSs and downlink users to maximize the signal-to-clutter-plus-noise ratio of sensing.  In addition, Cheng \emph{et al}. in \cite{Cheng2025} proposed a joint transmit beamforming and UAV trajectory optimization framework, in which multiple BSs communicate with multiple UAVs and simultaneously sensing a targeted three-dimensional (3D) space. The works in \cite{Zhang2025} and \cite{Cheng2025} assumed each user is served by a single BS, which constrains the potential for the cooperation among the BSs in serving communication users.

 It is worth noting that cellular ISAC systems still cannot overcome the adverse effects of line-of-sight (LOS) blockage between BSs and low-altitude targets below 100m in urban scenarios. To address this problem, deploying a reconfigurable intelligent surface (RIS) on buildings is an innovative solution to adjust wireless environment \cite{Chen2024}. In \cite{Liao2023}, Liao \emph{et al}. proposed a jointly BS's beamformers and RIS's phase shifting matrix optimization strategy for a RIS-assisted ISAC system, where a BS senses multi-target and communicates with multi-user. Moreover, the scenario is expanded to multiple BSs with co-located full-duplex transceivers in \cite{Yang2024}. However, the self-interference between transceiver antennas will harm system performance seriously and the authors assumed multiple targets positioned at certain locations \cite{Niu2025}.

Different from the above-mentioned works, this paper studies a RIS-assisted coordinated multi-point (CoMP) ISAC network, which is designed to provide multi-user communication and low-altitude sensing coverage for the prescribed sensing region. Our objective is to minimize the total transmit power through the joint optimization of transmit beamforming vectors and RIS phase shifts parameters under the constraints of sensing signal-to-noise ratio (SNR) and communication spectral efficiency (SE) requirements. Then, an alternating optimization (AO) algorithm is proposed to solve the problem. Specifically, we alternately optimize the beamforming and RIS via using semi-definite relaxation (SDR) technique. Numerical results are provided to show the proposed strategy significantly reduces the total transmit power compared to benchmarks while also achieving continuous sensing coverage.


\section{System Model}
\begin{figure}
\centering
\includegraphics[scale=0.45]{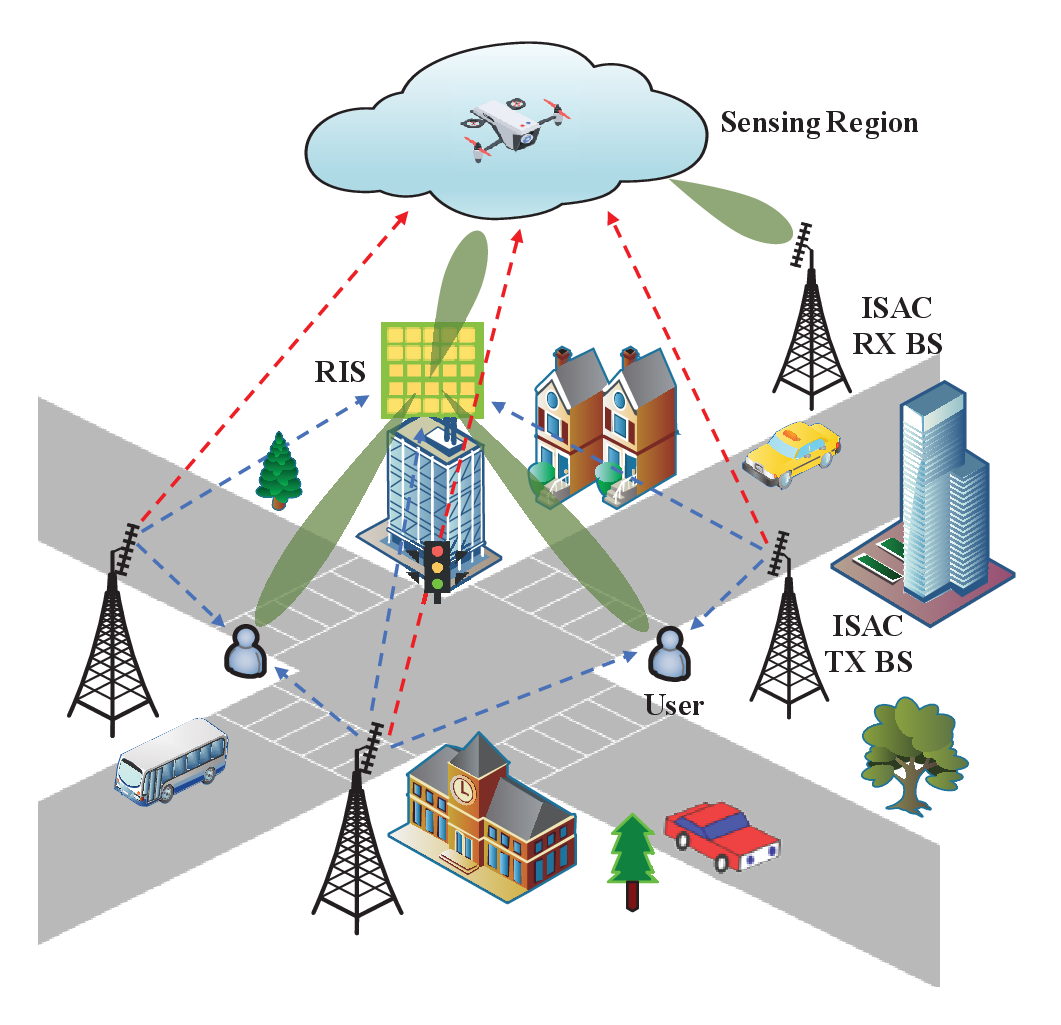}
\caption{\; Illustration of RIS-assisted CoMP ISAC networks}\label{network}
\vspace{-0.5cm} 
\end{figure}

As shown in Fig. \ref{network},  we consider a RIS-assisted CoMP ISAC system comprising $K$ downlink single-antenna users, a RIS, $J$  ISAC transmit (TX) BSs and one ISAC receive (RX) BS, which are connected by a central processor. The system needs to serve ground communication users, while sensing the unauthorized targets over the prescribed region ${\mathcal O}$  located at $h$ meters above the ground. The BSs and RIS are equipped with an uniform planar array (UPA) of $N = {N_x} \times {N_y}$ antennas and $M = {M_x} \times {M_y}$ reflecting elements, where $(\cdot)_{x}$  and $(\cdot)_{y}$  denote the number of elements along x and y axes, respectively. The set of users and TX BSs are denoted as $ \mathcal{K}=\{1, 2, ... ,K\}$  and $\mathcal{J}=\{1, 2, ... ,J\}$. Without loss of generality, we consider a 3D Cartesian coordinate system. Let $\mathbf{q}_j=[x_j,y_j,H_j]^\text{T}, \ j\in\mathcal{J}$, $\mathbf{q}_0=[x_0,y_0,H_0]^\text{T}$, $\mathbf{q}=[x,y,h]^\text{T}$ and $\mathbf{u}_k=[x_k,y_k,0]^\text{T}, \ k\in\mathcal{K}$, denote the positions of TX BS $j$, RX BS, RIS, and user $k$, respectively, where $H_j$  and $H_0$ are the height of TX BS $j$ and RX BS. Moreover, let $\mathbf{o}\in {\mathcal O}$  denotes the location of an arbitrary target within the prescribed sensing coverage region.

The coordinated transmit signal by BS $j$ can be expressed as
\begin{equation}
{{\mathbf{s}}_j} = \sum\limits_{k = 1}^K {{{\bm{\omega }}_{j,k}}{x_k}}  + {{\bm{\omega }}_{j,0}}{x_0},
\end{equation}
where ${x_k}$  and ${x_0}$  denote the transmitted data for user $k$ and the dedicated sensing signal, ${{\bm{\omega }}_{j,k}} \in \mathbb{C}^{N \times 1}$ and ${{\bm{\omega }}_{j,0}} \in \mathbb{C}^{N \times 1}$  represent the beamforming vectors, respectively.
\subsection{Communication Model}
We consider the users have dedicated hardware and modules to cancellate sensing interference. Specifically, given that the sensing signal ${x_0}$ is known \emph{a priori} to all users, each user can reconstruct and subsequently subtract the sensing interference based on the channel state information \cite{Cheng2024}. Therefore, the resulting communication signal at the user $k$, comprising both the BS-user link and BS-RIS-user link, can be expressed as
\begin{equation}
{y_k} = \underbrace {\sum\limits_{j = 1}^J {{\bf{g}}_{j,k}^{\rm{H}}{{\bm{\omega }}_{j,k}}{x_k}} }_{{\text{desired signals}}} + \underbrace {\sum\limits_{j = 1}^J {\sum\limits_{i = 1,i \ne k}^K {{\bf{g}}_{j,k}^{\rm{H}}{{\bm{\omega }}_{j,i}}{x_i}} } }_{{\text{interference signals}}} + {n_k},
\end{equation}
where ${{\bf{g}}_{j,k}} = {{\bf{h}}_{j,k}} + {\mathbf{G}}_{{\text{R}},j}^{\rm{H}}{{\mathbf{\Phi }}^{\rm{H}}}{{\bf{h}}_{{\text{R}},k}} \in  {\mathbb{C}^{N \times 1}}$ denotes the effective channel vector from BS $j$ to user $k$, and ${n_k}$ denotes the additive white Gaussian noise (AWGN) with power $\sigma _k^2$. ${\mathbf{\Phi }} = {\text{diag}}\left( {{e^{j{\phi _1}}}, \cdots ,{e^{j{\phi _M}}}} \right) \in {\mathbb{C}^{M \times M}}$  denotes as the diagonal phase shifts matrix of the RIS, where $\phi _m  \in \left[ {0,2\pi } \right)$  is the phase shift of the  $m$-th element of the RIS. ${{\mathbf{G}}_{{\text{R}},j}} \in {\mathbb{C}^{M \times N}}$, ${{\mathbf{h}}_{{\text{R}},k}} \in {\mathbb{C}^{M \times 1}}$  and ${{\mathbf{h}}_{j,k}} \in {\mathbb{C}^{N \times 1}}$  denote the channels from BS $j$ to the RIS, from the RIS to user $k$, and from BS $j$ to user $k$, where all user-associated links are assumed to follow Rician fading. Thus, the channel between BS $j$ and user $k$ is given by
\begin{equation}
{{\mathbf{h}}_{j,k}} = \sqrt {\frac{\zeta }{{\left\| {{{\mathbf{u}}_k} - {{\mathbf{q}}_j}} \right\|_2^\alpha }}} \left( {\sqrt {\frac{1}{{\beta  + 1}}} {\mathbf{h}}_{j,k}^{{\text{NLOS}}} + \sqrt {\frac{\beta }{{\beta  + 1}}} {\mathbf{h}}_{j,k}^{{\text{LOS}}}} \right),
\end{equation}
where $\zeta$  is the path loss at the reference distance of 1m, $\alpha$  is the path loss exponent, $\beta$ is the Rician factor, ${\mathbf{h}}_{j,k}^{{\text{NLOS}}}$  is the  non-line-of-sight (NLOS) component, with each of its elements following ${\mathcal{CN}}(0,1)$,  and  ${\mathbf{h}}_{j,k}^{{\text{LOS}}} = {{\bm{\alpha }}_{\text{B}}}\left( {{\theta _{j,k}},{\varphi _{j,k}}} \right){e^{ - j2\pi {f_c}{\tau _{j,k}}}}$ is the LOS component, where $f_c$  is the carrier frequency,  ${\tau _{j,k}}$ is the propagation delay from BS $j$ to user $k$, ${\theta _{j,k}}$ and ${\varphi _{j,k}}$  are the azimuth and elevation angles of user $k$ with respect to BS $j$, respectively, and ${{\bm{\alpha }}_{\text{B}}}\left( {{\theta _{j,k}},{\varphi _{j,k}}} \right)$  is the transmit array response vector, defined as
\begin{equation}
{{\bm{\alpha }}_{\text{B}}}\left( {{\theta _{j,k}},{\varphi _{j,k}}} \right) = {{\bm{\alpha }}_{{\text{B}},x}} \otimes {{\bm{\alpha }}_{{\text{B}},y}},
\end{equation}
where ${{\bm{\alpha }}_{{\text{B}},x}} = {\left[ {1,{e^{ - j{{\bar d}_x}}},...,{e^{ - j{{\bar d}_x}\left( {{N_x} - 1} \right)}}} \right]^{\text{T}}}$, ${{\bm{\alpha }}_{{\text{B}},y}} = {\left[ {1,{e^{ - j{{\bar d}_y}}},...,{e^{ - j{{\bar d}_y}\left( {{N_y} - 1} \right)}}} \right]^{\text{T}}}$, ${\bar d_x} = \frac{{2\pi {d_x}}}{\lambda }\cos {\theta _{j,k}}\sin {\varphi _{j,k}}$, ${\bar d_y} = \frac{{2\pi {d_y}}}{\lambda }\sin {\theta _{j,k}}\sin {\varphi _{j,k}}$, where $d_x$  and $d_y$  denote the inter-element spacing in $x$ and $y$ axes, respectively, $\lambda$ is the wavelength, and  $\otimes$   denotes the Kronecker product. While the channel between TX BSs and RIS is assumed to be an LOS path, ${{\bf{G}}_{{\text{R}},j}} = \sqrt {\frac{\kappa }{{\left\| {{\bf{q}} - {{\bf{q}}_j}} \right\|_2^\alpha }}} {{\bm{\alpha }}_{\text{R}}}\left( {{\theta _{{\text{R}},j}},{\varphi _{{\text{R}},j}}} \right){\bm{\alpha }}_{\text{B}}^{\text{H}}\left( {{\psi _{{\text{R}},j}},{\eta _{{\text{R}},j}}} \right){e^{ - j2\pi {f_c}{\tau _{{\text{R}},j}}}}$. Therefore, the spectral efficiency of user $k$ can be expressed as
\begin{equation}
{R_k} = {\log _2}\left( {1 + \frac{{\left\| {\sum\limits_{j = 1}^J {{\bf{g}}_{j,k}^{\rm{H}}{{\bm{\omega }}_{j,k}}} } \right\|_2^2}}{{\left\| {\sum\limits_{j = 1}^J {\sum\limits_{i = 1,i \ne k}^K {{\bf{g}}_{j,k}^{\rm{H}}{{\bm{\omega }}_{j,i}}} } } \right\|_2^2 + \sigma _k^2}}} \right).
\end{equation}
\subsection{Sensing Model}
 To achieve the best sensing performance, both the data and sensing signals are used for sensing. Meanwhile, to avoid self-interference, we employ the sensing RX BS to monitor the unauthorized targets in the prescribed sensing region. The RX BS performs sensing by receiving the echo signals from two distinct propagation paths. One is the direct path that the signal transmitted by TX BSs is reflected directly by targets to the RX BS. Another is the RIS-assisted path that the signal from TX BSs is first reflected by the RIS toward targets, and then reflected by targets to the RX BS. We assume that the target-free channel between TX BSs and RX BS is acquired prior to sensing and the transmit signals are also known at the RX BS. Therefore, the target-free part of the received signal can be cancelled at the RX BS. Hence, the echo signal received by the RX BS is given by
\begin{equation}
{\bf{y}} = \sum\limits_{j = 1}^J {\sum\limits_{k = 0}^K {{{\bf{H}}_{0,j}}{{\bm{\omega }}_{j,k}}{x_k}} }  + {\bf{n}},
\end{equation}
where ${{\mathbf{H}}_{0,j}} = {{\mathbf{h}}_{0,o}}\left( {{\mathbf{h}}_{j,o}^{\text{H}} + {\mathbf{h}}_{{\text{R}},o}^{\text{H}}{\mathbf{\Phi }}{{\mathbf{G}}_{{\text{R}},j}}} \right) \in {\mathbb{C}^{N \times N}}$  with ${\mathbf{h}}_{0,o} \in \mathbb{C}^{N \times 1}$,  ${\mathbf{h}}_{j,o} \in \mathbb{C}^{N \times 1}$ and ${\mathbf{h}}_{{\text{R}},o}\in \mathbb{C}^{M \times 1}$  denoting the channel vectors from the target to RX BS, TX BS $j$ to target, and the RIS to target, respectively. The channel between the RIS and target is assumed to be an LOS path, which can be  modeled as
\begin{equation}
{{\bf{h}}_{{\text{R}},o}} = \sqrt {\frac{\kappa }{{\left\| {{\bf{o}} - {\bf{q}}} \right\|_2^\alpha }}{\sigma _{{\text{rcs}}}}} {{\bm{\alpha }}_{\text{R}}}\left( {{\theta _{{\text{R}},o}},{\varphi _{{\text{R}},o}}} \right){e^{ - j2\pi {f_c}\tau_{{\text{R}},o} }},
\end{equation}
where ${\sigma _{{\text{rcs}}}} $ is the radar cross section (RCS) of the target in the prescribed sensing region, ${{\bm{\alpha }}_{\text{R}}}\left( {{\theta _{{\text{R}},o}},{\varphi _{{\text{R}},o}}} \right)$ is the steering vectors at direction ${\theta _{{\text{R}},o}}$ and ${\varphi _{{\text{R}},o}}$, and $\tau_{{\text{R}},o}$ is the propagation delay. Since the sensing tasks for detection and localization depend on the quality of the received signal, we employ the sensing SNR to evaluate the sensing coverage performance \cite{Zhang2016,Zhang2025b}. Hence, the sensing SNR for a target located at $\mathbf{o}\in {\mathcal O}$   can be expressed as
\begin{equation}
\gamma \left( {\bf{o}} \right) = \frac{{\left\| {\sum\limits_{j = 1}^J {\sum\limits_{k = 0}^K {{{\bf{H}}_{0,j}}{{\bm{\omega }}_{j,k}}} } } \right\|_2^2}}{{{\sigma ^2}}},
\end{equation}
where ${\sigma ^2}$ denotes the power of the AWGN.

\section{Problem Formulation and Solution}
This paper intends to minimize the total transmit power of  BSs by jointly optimizing the transmit beamforming vectors and the RIS phase shifts matrix, while guaranteeing the minimum SE required by communication users, and the minimum SNR required for sensing the prescribed sensing coverage region $\mathcal{O}$. To make the optimization problem tractable, $\mathcal{O}$  is approximated by a set of discretized points $\left\{ {{{\bf{o}}_l}} \right\}_{l = 1}^L$, where $L = {L_x} \times {L_y}$  with $L_x$  and  $L_y$ denoting the number of grid points along each dimension. Therefore, the considered problem is formulated as
\begin{subequations}
\begin{align}
\Pact_1: \;
\min_{ \{{\bm{\omega }}_{j,k}\}, \{{\bm{\omega }}_{j,0}\}, {\mathbf{\Phi }}}  \quad &\ \sum\limits_{j = 1}^J {\sum\limits_{k = 0}^K {\left\| {{{\bm{\omega }}_{j,k}}} \right\|_2^2} } \\
\text{s.t.} \quad &\
    {R_k} \ge {R_{{\text{req}}}},     \\
    \qquad& \ \gamma \left( {{{\bf{o}}_l}} \right) \ge {\gamma _{{\text{req}}}},  \\
    \qquad& \ {\left\| {{e^{j{\phi _m}}}} \right\|_2} = 1, \label{eq_P1d}
\end{align}
\end{subequations}
where  ${R_{{\text{req}}}}$ denotes the minimum SE of user $k$,  ${\gamma _{{\text{req}}}}$ is the predefined sensing SNR threshold. Constraint (\ref{eq_P1d}) is the unit-modulus requirements of RIS phase shifts. The problem $\mathscr{P}_1$  is difficult to solve in general due to the coupled optimization variables $\{{\bm{\omega }}_{j,k}\}$, $\{{\bm{\omega }}_{j,0}\}$ and  ${\mathbf{\Phi }}$. Therefore, we first split problem $\mathscr{P}_1$  into two sub-problems, and develop an algorithm to address them in an alternating manner.

\subsection{Optimal BSs Transmit Beamforming}
Let  ${{\bm{\omega }}_k} = {\left[ {{\bm{\omega }}_{1,k}^{\text{T}}, \cdots ,{\bm{\omega }}_{J,k}^{\text{T}}} \right]^{\text{T}}} \in {\mathbb{C}^{NJ \times 1}}$,  ${{\bf{g}}_k} = {\left[ {{\bf{g}}_{1,k}^{\text{T}}, \cdots ,{\bf{g}}_{J,k}^{\text{T}}} \right]^{\text{T}}} \in {\mathbb{C}^{NJ \times 1}}$, and ${{\bf{H}}_0} = \left[ {{{\bf{H}}_{0,1}}, \cdots ,{{\bf{H}}_{0,J}}} \right] \in {\mathbb{C}^{N \times NJ}}$. With fixed the RIS phase shifts, the transmit beamforming sub-problem is expressed as
\begin{subequations}
\begin{align}
\Pact_2: \;
\min_{ \{{{\bm{\omega }}_{k}} \} }  \quad &\ \sum\limits_{k = 0}^K {{\bm{\omega }}_k^{\text{H}}{{\bm{\omega }}_k}}  \\
 \text{s.t.} \quad &\
    \frac{{\left\| {{\bf{g}}_k^{\text{H}}{{\bm{\omega }}_k}} \right\|_2^2}}{{\sum\limits_{i = 1,i \ne k}^K {\left\| {{\bf{g}}_k^{\text{H}}{{\bm{\omega }}_i}} \right\|_2^2 + } \sigma _k^2}} \ge {2^{{R_{{\rm{req}}}}}} - 1{\rm{ }},  \label{eq_P2b}    \\
    \qquad& \ \left\| {\sum\limits_{k = 0}^K {{{\bf{H}}_0}{{\bm{\omega }}_k}} } \right\|_2^2 \ge {\sigma ^2}{\gamma _{{\rm{req}}}}.  \label{eq_P2c}
\end{align}
\end{subequations}
However, problem $\mathscr{P}_2$   is still non-convex due to the constraint (\ref{eq_P2b}) and (\ref{eq_P2c}). Defining  ${{\bf{G}}_k} = {{\bf{g}}_k}{\bf{g}}_k^{\text{H}} \in {\mathbb{C}^{NJ \times NJ}}$ and  ${{\bf{W}}_k} = {{\bm{\omega }}_k}{\bm{\omega }}_k^{\text{H}} \in {\mathbb{C}^{NJ \times NJ}}$,  problem  $\mathscr{P}_2$  can be reformulated as
\begin{subequations}
\begin{align}
\Pact_{2.1}: \;
\min_{ \{{{\mathbf{W}}_k}\} }  \quad &\ \sum\limits_{k = 0}^K {{\text{tr}}\left( {{{\mathbf{W}}_k}} \right)}   \\
 \text{s.t.} \quad &\
    {\text{tr}}\left( {{{\bf{G}}_k}{{\bf{W}}_k}} \right) \ge S \left( {\sum\limits_{i = 1,i \ne k}^K {{\text{tr}}\left( {{{\bf{G}}_k}{{\bf{W}}_i}} \right) + } \sigma _k^2} \right),     \\
    \qquad& \  {\text{tr}}\left( {{{\bf{H}}_0}\left( {\sum\limits_{k = 0}^K {{{\bf{W}}_k}} } \right){\bf{H}}_0^{\text{H}}} \right) \ge {\sigma ^2}{\gamma _{{\rm{req}}}},    \\
    \qquad& \  {\text{rank}}\left( {{{\bf{W}}_k}} \right)=1,  \label{eq_P21d} \\
    \qquad& \  {{\bf{W}}_k}  \succeq 0,
\end{align}
\end{subequations}
where $S={{2^{{R_{{\rm{req}}}}}} - 1}$. Problem  $\mathscr{P}_{2.1}$  is still non-convex due to the rank-one constraint (\ref{eq_P21d}). Thus, we adopt the SDR technique to drop them, and hence problem  $\mathscr{P}_{2.1}$ is a semi-definite program (SDP) problem. We can use the convex optimization solvers to obtain the optimal solution $\left\{ {{\bf{W}}_k^ \star } \right\}$. Moreover, the obtained solution is bound to satisfy  ${\text{rank}}\left( {{\bf{W}}_k^\star} \right)= 1$, which can be mathematically proved \cite{Huang2010,Li2022}.  Therefore, the optimal solution $\left\{ {{\bm{\omega }}_k^\star} \right\}$  can be obtained through eigenvalue decomposition (EVD) method. Let ${\lambda _k}$  denotes the maximal eigenvalue of  ${\bf{W}}_k^ \star $, and ${{\bm{\nu }}_k}$  denotes the corresponding eigenvector associated with  ${\lambda _k}$. Thus,  ${\bm{\omega }}_k^\star = \sqrt {{\lambda _k}} {{\bm{\nu }}_k}$.

\subsection{Optimal RIS Phase Shifts}
With the fixed transmit beamforming, the phase shifts optimization sub-problem can be transformed to a feasibility-check problem, which is formulated as
\begin{align}
\Pact_{3}: \;
 \text{Find} \quad &\ {\bf{\Phi }}  \notag  \\
 \text{s.t.} \quad &\
    (\ref{eq_P1d}), (\ref{eq_P2b}), (\ref{eq_P2c}). \notag
\end{align}
Here, all constraints in problem $\mathscr{P}_{3}$  are non-convex. We then reformulate the constraint (\ref{eq_P2b}) and (\ref{eq_P2c}) for the convenience of solving the problem. For constraint (\ref{eq_P2b}), by using ${\bf{h}}_{{\text{R}},k}^{\text{H}}{\bf{\Phi }}{{\bf{G}}_{{\text{R}},j}} = {{\bf{v}}^{\text{H}}}{{\bf{\Psi }}_{j,k}}$, where ${\bf{v}} = {\left[ {{e^{j{\phi _1}}}, \cdots ,{e^{j{\phi _M}}}} \right]^{\text{T}}} \in {\mathbb{C}^{M \times 1}}$, ${{\bf{\Psi }}_{j,k}} = {\text{diag}}\left( {{\bf{h}}_{{\text{R}},k}^{\text{H}}} \right){{\bf{G}}_{{\text{R}},j}} \in {\mathbb{C}^{M \times N}}$, defining ${{\bf{\Psi }}_k} = \left[ {{{\bf{\Psi }}_{1,k}}, \cdots ,{{\bf{\Psi }}_{J,k}}} \right] \in {\mathbb{C}^{M \times NJ}}$  and   ${{\bf{h}}_k} = {\left[ {{\bf{h}}_{1,k}^{\text{T}}, \cdots ,{\bf{h}}_{J,k}^{\text{T}}} \right]^{\text{T}}} \in {\mathbb{C}^{NJ \times 1}}$, (\ref{eq_P2b}) can be equivalently expressed as $\left\| {\left( {{{\bf{v}}^{\text{H}}}{{\bf{\Psi }}_k} + {\bf{h}}_k^{\text{H}}} \right){{\bm{\omega }}_k}} \right\|_2^2 \ge S \left( {\sum\limits_{i = 1,i \ne k}^K {\left\| {\left( {{{\bf{v}}^{\text{H}}}{{\bf{\Psi }}_k} + {\bf{h}}_k^{\text{H}}} \right){{\bm{\omega }}_i}} \right\|_2^2}  + \sigma _k^2} \right)$. Letting ${{\bf{C}}_k} = {\left[ {{\bf{\Psi }}_k^{\text{H}},{{\bf{h}}_k}} \right]^{\text{H}}} \in {\mathbb{C}^{\left( {M + 1} \right)\times NJ}}$ and ${\bf{\tilde v}} = {\left[ {{{\bf{v}}^{\text{H}}},1} \right]^{\text{H}}} \in {\mathbb{C}^{(M + 1)\times 1}}$, (\ref{eq_P2b}) can be converted into
\begin{equation}\label{eq_RISc}
{{\bf{\tilde v}}^{\text{H}}}{{\bf{F}}_k}{\bf{\tilde v}} \ge S \left( {\sum\limits_{i = 1,i \ne k}^K {{{{\bf{\tilde v}}}^{\text{H}}}{{\bf{F}}_i}{\bf{\tilde v}}}  + \sigma _k^2} \right),
\end{equation}
where  ${{\bf{F}}_k} = {{\bf{C}}_k}{{\bm{\omega }}_k}{\bm{\omega }}_k^{\text{H}}{\bf{C}}_k^{\text{H}}$, ${{\bf{F}}_i} = {{\bf{C}}_k}{{\bm{\omega }}_i}{\bm{\omega }}_i^{\text{H}}{\bf{C}}_k^{\text{H}}$.

For constraint (\ref{eq_P2c}), by using  ${\bf{h}}_{{\text{R}},o}^{\text{H}}{\bf{\Phi }}{{\bf{G}}_{{\text{R}},j}} = {{\bf{v}}^{\text{H}}}{{\bf{\Psi }}_{j,o}}$, where ${{\bf{\Psi }}_{j,o}} = {\text{diag}}\left\{ {{\bf{h}}_{{\text{R}},o}^{\text{H}}} \right\}{{\bf{G}}_{{\text{R}},j}} \in {\mathbb{C}^{M \times N}}$, defining   ${{\bf{\Psi }}_o} = \left[ {{{\bf{\Psi }}_{1,o}}, \cdots ,{{\bf{\Psi }}_{J,o}}} \right] \in {\mathbb{C}^{M \times NJ}}$, ${{\bf{h}}_o} = {\left[ {{\bf{h}}_{1,o}^{\text{T}}, \cdots ,{\bf{h}}_{J,o}^{\text{T}}} \right]^{\text{T}}} \in {\mathbb{C}^{NJ \times 1}}$, (\ref{eq_P2c}) can be equivalently written as  ${\text{tr}}\left( {{{\bf{E}}_0}\left( {\sum\limits_{k = 0}^K {{{\bf{W}}_k}} } \right){\bf{E}}_0^{\text{H}}} \right) \ge {\sigma ^2}{\gamma_{\text{req}}}$, where ${{\bf{E}}_0} = {{\bf{h}}_{0,o}}\left( {{{\bf{v}}^{\text{H}}}{{\bf{\Psi }}_o} + {\bf{h}}_o^{\text{H}}} \right) \in {\mathbb{C}^{N \times NJ}}$. By defining ${{\bf{B}}_o} = {\left[ {{\bf{\Psi }}_o^{\text{H}},{{\bf{h}}_o}} \right]^{\text{H}}} \in {\mathbb{C}^{(M + 1) \times NJ}}$,  $s = {\bf{h}}_{0,o}^{\text{H}}{{\bf{h}}_{0,o}}$, (\ref{eq_P2c})  can be converted into
\begin{equation}\label{eq_RISs}
{\text{tr}}\left( {s{{{\bf{\tilde v}}}^H}{{\bf{U}}_o}{\bf{\tilde v}}} \right) \ge {\sigma ^2}{\gamma _{{\text{req}}}},
\end{equation}
where ${{\bf{U}}_o} = {{\bf{B}}_o}\left( {\sum\limits_{k = 0}^K {{{\bf{W}}_k}} } \right){\bf{B}}_o^{\text{H}}$.

Let  ${\bf{V}} = {\bf{\tilde v}}{{\bf{\tilde v}}^{\text{H}}} \in {\mathbb{C}^{\left( {M + 1} \right) \times \left( {M + 1} \right)}}$, the RIS phase shifts optimization problem can be expressed as
\begin{subequations}
\begin{align}
\Pact_{3.1}: \;
\text{Find}  \quad &\ {\bf{V}}  \\
 \text{s.t.} \quad &\
    {\text{tr}}\left( {{{\bf{F}}_k}{\bf{V}}} \right) \ge S \left( {\sum\limits_{i = 1,i \ne k}^K {{\text{tr}}\left( {{{\bf{F}}_i}{\bf{V}}} \right)}  + \sigma _k^2} \right),     \\
    \qquad& \ {\text{ tr}}\left( {s{{\bf{U}}_o}{\bf{V}}} \right) \ge {\sigma ^2}{\gamma _{{\text{req}}}}, \\
    \qquad& \  {\bf{V}} \succeq 0, {{\bf{V}}_{m,m}} = 1,   \\
    \qquad& \  {\text{rank}}\left( {\bf{V}} \right) = 1.  \label{eq_P31d}
\end{align}
\end{subequations}
\begin{algorithm} [t]
\caption{AO-Based Algorithm for Solving Problem $\mathscr{P}_1$ } 
\label{alg:algorithm1}
\hspace{-0.05in} {\bf Input:}    
 ${R_{{\text{req}}}} > 0$, ${\gamma_{{\text{req}}}} > 0$, threshold $\epsilon > 0$. \\
\hspace{0.02in} {\bf Output:}
Optimal transmit beamforming  $\{{\bm{\omega }}_{j,k}\}$,  $\{{\bm{\omega }}_{j,0}\}$, and RIS phase shifts  ${\mathbf{\Phi }}$. \\
\hspace{0.02in} {\bf Initialize:}
RIS phase shifts  $\mathbf{\Phi }^{(0)}$, and set the iteration number  $n = 1$.
\begin{algorithmic}[1]
\While{The fractional decrease of the objective value of problem $\mathscr{P}_1$  exceeds $\epsilon$ }
    \State Solve problem $\mathscr{P}_{2.1}$ for given  $\mathbf{\Phi }^{(0)}$, and obtain the solution  ${{\bm{\omega }}^{\left( n \right)}}$.
    \State Solve problem $\mathscr{P}_{3.1}$ for given  ${{\bm{\omega }}^{\left( n \right)}}$, and obtain the solution  ${{\bf{\tilde v}}^{\left( n \right)}}$. The updated  phase shifts matrix is  ${{\bf{\Phi }}^{\left( n \right)}} = {\text{diag}}\left( {{\mathbf{\tilde v}}_{1:M}^{\left( n \right)}} \right)$.
    \State Update  $n=n+1$.
\EndWhile
\end{algorithmic}
\end{algorithm}
Note that the unique non-convex constraint of problem $\mathscr{P}_{3.1}$ is the constraint (\ref{eq_P31d}). Therefore, we can relax the rank constraint (\ref{eq_P31d}), thus obtaining the optimal solution $ {{{\bf{V}}^ \star}}$  via standard convex solvers. If ${\text{rank}}\left( {{{\bf{V}}^\star}} \right)= 1$, we can obtain the optimal solution $ {{{{\bf{\tilde v}}}^\star}}$ by EVD. Otherwise, the Gaussian randomization can be used to derive rank-1 solutions. Since  ${\text{rank}}\left( {{{\bf{V}}^\star}} \right) \ne {\rm{1}}$, the eigenvalue decomposition of  $\bf{V}^\star$ can be expressed as  ${{\bf{V}}^\star} = {\bf{U\Sigma }}{{\bf{U}}^{\text{H}}}$, where ${\bf{U}} = \left[ {{{\bm{\nu }}_1}, \cdots ,{{\bm{\nu }}_{M + 1}}} \right]$ and ${\bf{\Sigma }} = {\text{diag}}\left( {{\lambda _1}, \cdots ,{\lambda _{M + 1}}} \right)$  are the identity matrix of the eigenvector and the diagonal matrix of the eigenvalue. The suboptimal solution to $\mathscr{P}_{3.1}$ is ${{\bf{\tilde v}}^\star} = {\bf{U}}{{\bf{\Sigma }}^{1/2}}{\bm{\chi }}$, where  ${\bm{\chi }} \sim \mathcal{CN}\left( {{\bf{0}},{{\bf{I}}_{M + 1}}} \right)$. With independently generated Gaussian random vectors  ${\bm{\chi }}$, the objective value of $\mathscr{P}_{1}$ is approximated as the maximum one attained by the best ${{\bf{\tilde v}}^\star}$ among all ${\bm{\chi }}$. Finally, the solution  ${\bf{v}}$ can be recovered by ${\bf{v}} = {e^{j \arg \left( {{{\left[ {\frac{{{{{\bf{\tilde v}}}^\star}}}{{{\bf{\tilde v}}_{\left[ {M + 1} \right]}^\star}}} \right]}_{1:M}}} \right)}}$. We summarize the overall AO algorithm in  Algorithm \ref{alg:algorithm1}.

\section{Numerical Results}
\subsection{Network Settings}
\begin{table}[t]
\centering
\setlength{\tabcolsep}{5mm}{
\caption{ Primary Simulation Parameters }
\label{tab1}
\begin{tabular}{lll}
   \toprule
   Symbol & Parameter & Value \\
   \midrule
   $J$  &Number of TX BSs   & 3  \\
   $K$  &Number of users   & 10  \\
   $N$  &Number of BS antennas   & 32  \\
   $M$  &Number of RIS elements   & 64  \\
   $f_c$ &Carrier frequency (GHz) & 3.5  \\
   $B$   &Signal bandwidth (MHz)   & 100  \\
   $N_0$ &Noise power density (dBm/Hz)& -174  \\
   $\zeta$  &channel power at 1m (dB)& -43 \\
   $\sigma_\text{rcs}$ &RCS of targets ($\text{m}^2$) &2 \\
   $L$  &Sensing sampling points & 16  \\
   \bottomrule
\end{tabular}}
\end{table}
In this section, we conduct a comprehensive performance evaluation of the proposed joint beamforming and reflection design for supporting low-altitude sensing coverage. Specifically, we consider a RIS-assisted CoMP ISAC network where the RIS is located at $(65\text{m},95\text{m},40\text{m})$, and the TX BSs and RX BS are centered at $(0\text{m},0\text{m}\,30\text{m})$, $(0\text{m},200\text{m},30\text{m})$, $(173\text{m},100\text{m},30\text{m})$, and $(58\text{m},100\text{m},40\text{m})$, respectively. The users are uniformly distributed in the network, excluding an inner circle of 50m around each TX BS. The sensing region is a rectangular region with horizontal demensions of $30\text{m} \times 30\text{m}$ and a vertical height of $60\text{m}$. The top view of the scenario is shown in Fig. \ref{fig_BF}. The path loss exponents are set to 3.7 and 2.8 for the TX BS-user and TX BS-target links, respectively, owing to the extensive obstacles and scatterers.  All other links adopt an exponent of 2. For small-scale fading, the RIS-target and RIS-TX BS channels are modeled as purely LOS with a Rician factor of $\infty$, whereas the remaining channels are characterized by a Rician factor of 3 dB. Other primary simulation parameters are listed in Table  \ref{tab1}.

\subsection{Performance Analysis}
\begin{figure}[t]
\centering
\includegraphics[scale=0.5]{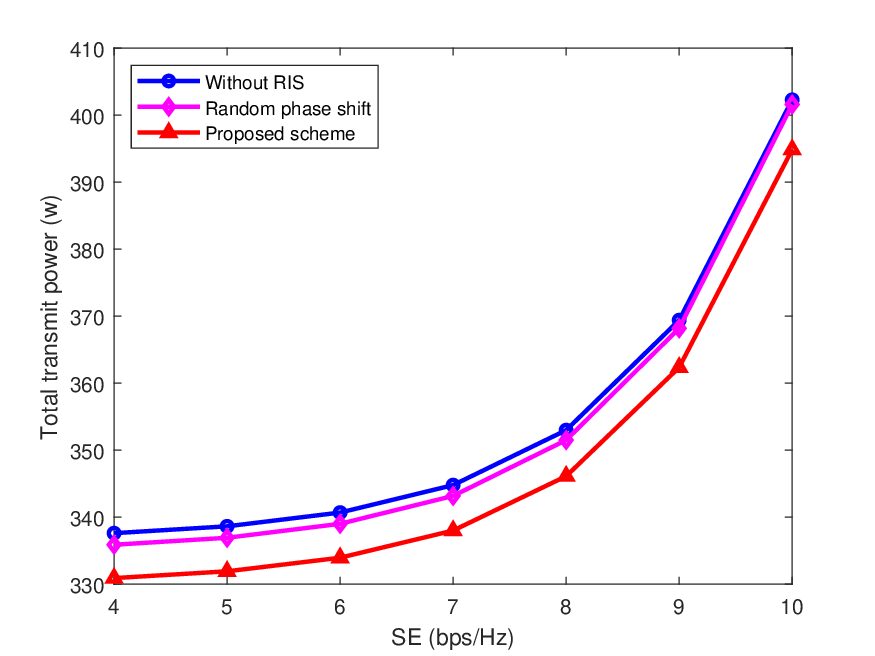}
\caption{\; Total transmit power versus communication SE for different schemes.}\label{fig_Power_SE}
\end{figure}

\begin{figure}[t]
\centering
\includegraphics[scale=0.5]{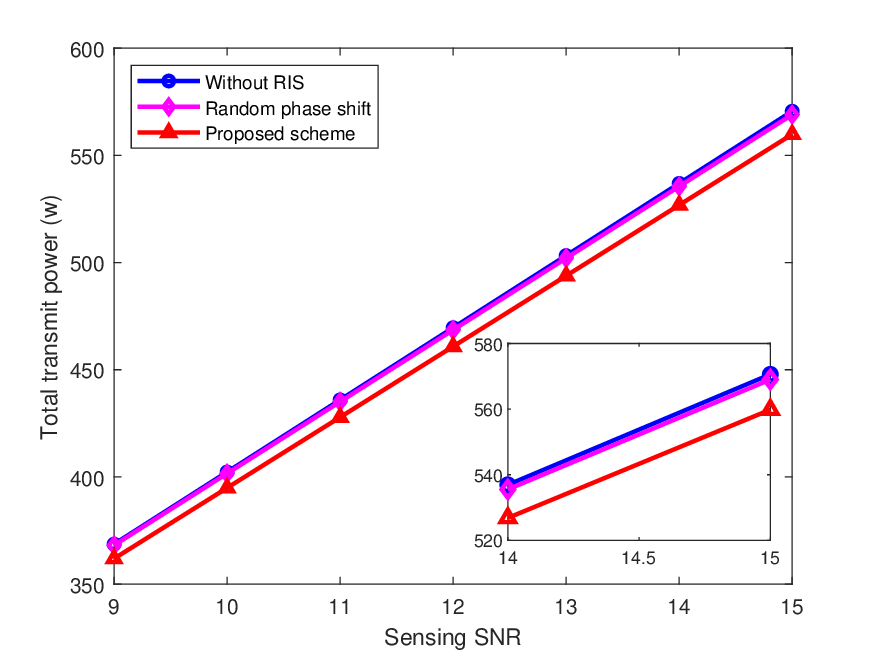}
\caption{\; Total transmit power versus sensing SNR for different schemes.}\label{fig_Power_SNR}
\end{figure}

To highlight the advantages of our proposed scheme for low-altitude sensing coverage, we compare it against two baseline schemes: 1) the optimal beamforming without RIS \cite{Li2024}. 2) the optimal beamforming with random phase shifts. Fig. \ref{fig_Power_SE} shows the total transmit power required by all the above schemes versus SE when $\gamma_{\text{req}}$=10 ${\text{dB}}$. It is observed that, by adding the RIS, the total transmit power can be reduced as compared to the case without RIS . This is because the RIS can help create a link to effectively illuminate the desired directions. Moreover, compare to all baseline schemes, the proposed scheme achieves the same SE with a significantly lower total transmit power. Specifically, when the requirement of SE is 10 bps/Hz, the required total transmit power is reduced by 7.42w and 6.73w relative to the case without RIS and the case with random phase shifts, respectively. The above results validate the importance of our joint beamforming and reflection design in energy conservation.

The impact of the sensing SNR requirements on total transmit power is depicted in Fig. \ref{fig_Power_SNR}, where $R_{\text{req}}$=10 ${\text{bps/Hz}}$ . When the value of sensing SNR is 15, the proposed scheme can reduce the power 10.45w and 9.19w compared to the without RIS and random phase shift schemes. The main reason is that the RIS optimizes its phase shifts to enhance the transmitted signal, thereby reducing the power requirement. Moreover, as the SNR increases, the required total transmit power  exhibits  a linear growth trend. Consequently, the deployment of a RIS with the optimized phase adjustment emerges as a promising strategy for enabling energy-efficient design in low-altitude sensing coverage systems.

\begin{figure}[t]
\centering
\includegraphics[scale=0.5]{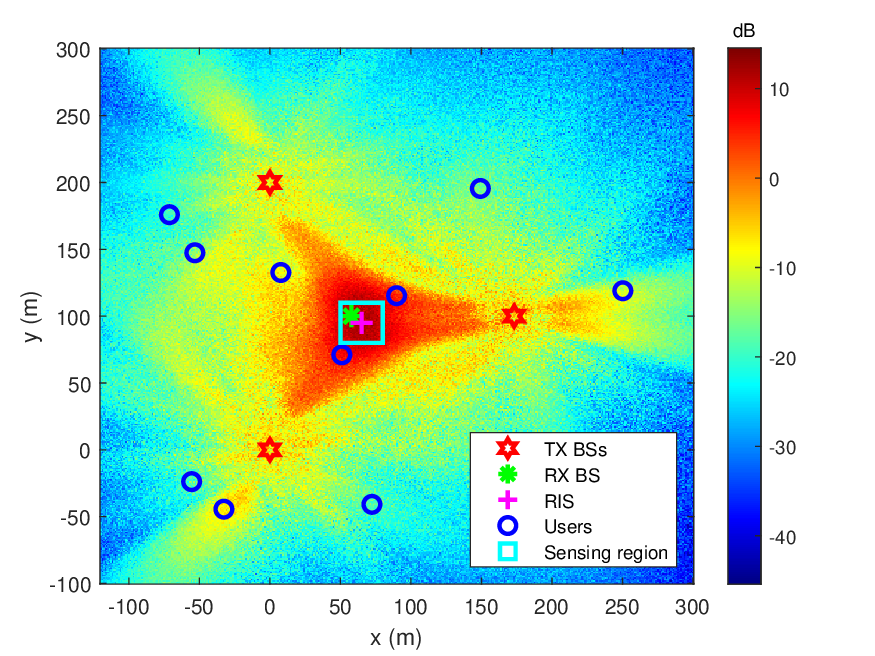}
\caption{\; Spatial distribution of the sensing SNR at 60m altitude using the proposed scheme.}\label{fig_BF}
\end{figure}
Fig. \ref{fig_BF} illustrates the spatial distribution of the sensing SNR  at  60m altitude achieved by our proposed design, under the communication SE and sensing SNR requirements of  $R_{\text{req}}$=10 ${\text{bps/Hz}}$ and $\gamma_{\text{req}}$=10 ${\text{dB}}$. As we can see that the  prescribed sensing region is entirely enclosed within the high-SNR zone, where the sensing SNR is above 10 $ {\text{dB}}$, satisfying the sensing coverage requirement. Moreover, the high-SNR coverage extends contiguously beyond the prescribed region and the sensing SNR values exceed 0  ${\text{dB}}$ within the central region encompassed by the TX BSs. This is because the cooperation among TX BSs can enhance the transmit gain through coherent joint transmission, while the optimized RIS phase shifts effectively direct signal toward the targeted area, thereby improving sensing coverage performance.

\begin{figure}[t]
\centering
\includegraphics[scale=0.5]{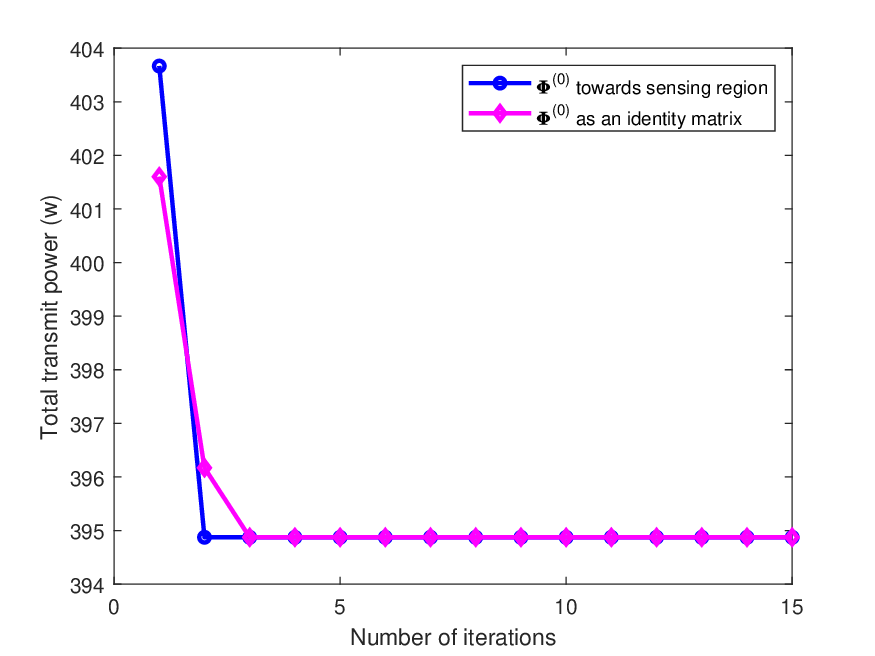}
\caption{\; The convergence behavior of the proposed Algorithm \ref{alg:algorithm1}.}\label{fig_convergence}
\end{figure}

Fig. \ref{fig_convergence} shows the convergence behavior of the proposed Algorithm \ref{alg:algorithm1} with different initial RIS phase shifts matrix $\mathbf{\Phi }^{(0)}$, where $R_{\text{req}}$=10 ${\text{bps/Hz}}$ and $\gamma_{\text{req}}$=10 ${\text{dB}}$. It is observed that the total transmit power obtained by our proposed algorithm is quickly decreased with the number of iterations. Moreover, both initial phase shift configurations converge to the same total transmit power after only about 3 iterations, confirming the low complexity of the proposed algorithm.

\section{Conclusion}
In this paper, we considered a RIS-assisted CoMP ISAC system, where the RIS is deployed to enable simultaneous communication with multiple ground users and sensing the prescribed   low-altitude region. We proposed the joint BS's transmit beamforming and RIS's phase shift design to minimize the total transmit power under constrains for SE, RIS phase shifts, and sensing SNR. To address the formulated non-convex optimization problem, we  developed a low-complexity algorithm based on the AO and SDR techniques. Further, we considered cases like without RIS and random phase shifts as the baseline schemes. Numerical results demonstrate that the proposed scheme achieves significant gains in energy efficiency over baseline schemes. Furthermore, it ensures continuous sensing coverage while maintaining a high SNR with low computational complexity. These results provide valuable  insights  for designing low-altitude sensing coverage systems.




\section*{Acknowledgment}
This paper was supported by the Natural Science Foundation of China under Grant No. 62171160, the Guangdong Provincial Key Laboratory (2024) (No. 2024KSYS023) and also Major Key Project of PCL (PCL2024A01).



%
\bibliographystyle{IEEETrans}
\bibliography{coverage}

\end{document}